\documentclass[conference,10pt,a4]{IEEEtran}
\usepackage{pslatex}
\usepackage{graphicx}
\usepackage{pst-node}
\usepackage{textcmds}
\usepackage{textcmds}

\onecolumn
\begin{document}

\title{\Large Service Oriented Architecture in Network Security - \\a novel Organisation in Security Systems}

\author{
\authorblockN{Michael Hilker \& Christoph Schommer}
\authorblockA{{\small University of Luxembourg, Campus Kirchberg}\\
{\small Dept. of Computer Science and Communication}\\
{\small 6, Rue Richard Coudenhove-Kalergi, L-1359 Luxembourg}\\
{\small Email: \{michael.hilker, christoph.schommer\}@uni.lu}}}

\maketitle

\thispagestyle{plain}

\begin{abstract}
\noindent Current network security systems are a collection of various security components, which are directly installed in the operating system. These check the whole node for suspicious behaviour. Armouring intrusions e.g. have the ability to hide themselves from being checked. We present in this paper an alternative organisation of security systems. The node is completely virtualized with current virtualization systems so that the operating system with applications and the security system is distinguished. The security system then checks the node from outside and the right security components are provided through a service oriented architecture. Due to the running in a virtual machine, the infected nodes can be halted, duplicated, and moved to other nodes for further analysis and legal aspects. This organisation is in this article analysed and a preliminary implementation showing promising results are discussed
\end{abstract}

\begin{quote}
\end{quote}

\IEEEpeerreviewmaketitle

\section{Introduction}
\label{secIntroduction}
In network security, well known security systems protect a computer network against various types of electronical attacks. The attacks are e.g. worms and viruses but also hacker or internal attacks performed by the normal users of the network. The security systems are a collection of various security components as antivirus software, firewalls, and intrusion detection systems \cite{Szo05}. These are directly in the operating system installed and observe the nodes for suspicious behaviour. The components lack from cooperative workflows in order to correlates events for abnormal behaviour detection \cite{Gartner07}. Also redundant checks lead to an increase in the required resources. This organisation of security components is a major weakness especially in coping with upcoming intrusions \cite{Szo01}. E.g. the bradeley virus is computational hard to detect when such an organisation is used \cite{Fil05}.

We concern with the organisation of security components and introduce a more sophisticated way. The whole node is virtualized using an virtualization system as VMware. The operating and security systems run in a node in different virtual machines so that the security components check the node from outside. The features of current virtualization systems are used with the ability to halt, duplicate, and move virtual machines. This leads to a different handling of infections because the infected virtual machine is duplicated and saved for further analysis and legal aspects. A service oriented architecture (SOA) provides on demand the right security components, which is implemented through the exchange or adding of virtual machines. We concern the advantages and disadvantages of such an environment and conclude with the current project status.

\section{Architecture}
Each node contains an architecture organised in four layers: the first layer is the hardware and the second layer is a core operating system providing the kernel to access the hardware and the virtualization system to run several virtual machines simultaneously. The third layer contains the virtual machines with different operating systems and one virtual machine for security. This implements the security environment, which ensures the access to all other virtual machines for scanning purposes. The fourth layer is the application layer with user's applications and installed security components, which are connected to the security environment. 

In the network, one or more security servers exist, which provide the right virtual machine with installed security environment and security components. This server provides the virtual machine to new nodes and ensures that each node is properly secured. Furthermore, it contains more analysis systems to scan virtual machines deeply. With this implementation, the security components are seen as different services provided by the security server and distributed to the node where they are required. This is expanded to a service oriented architecture where the right security components are provided on demand according to the current situation in the network. Especially with the novel more and more different node types as mobile handhelds connected to the network and thin clients, the required security components in a node has to be adapted due to different available resources and security issues. With the introduced architecture, the security components can be easily adapted to the security requirements of the node.

The maintenance of security systems changes according to the architecture. If a new node connects to the network, the security server checks the node if the right architecture is installed and provides an virtual machine with the security environment and installed security components. Through integrating this workflow in the DHCP, a network is properly secured because only when the node has a running security system it receives access to the network. New security components or changes in the required components in a node are quickly resolved: the security server provides a new virtual machine that exchanges the current machines on the node. 

\section{Implementation Issues}
The implementation of the architecture is feasible. Current virtualization systems as VMware or KVM provide most of the required features. VMware also provides mainboards where the layer one and two is directly installed. The only missing feature is the ability that the security components of the security environment are able to access the other virtual machines for scanning purposes. However, this can be implemented through an extension of the virtualization system.  

With the right implementation, well known security components - e.g. antivirus software, firewall, and intrusion detection system - are facilitated. These are installed in the security environment and a guard measures the required data from the operating system and presents it to the security component. This analyses the data and defines the response, which is executed by the guard accordingly. Consequently, all existing security components are reusable.

This architecture provides an improvement in the implementation of security components. These are platform independent running in the security environment and must consequently not adapted when the used operating system or used hardware platform changes. In addition, the security environment gathers the data to analyse and perform the response accordingly, which must not be implemented in the security component. This leads to a faster deployment of novel approaches.

The project status is that the architecture of the nodes and of the network is designed and theoretically analysed. Different proof-of-concept implementations of the virtualization system have been realised to analyse the features of these. Various parts of the implementation are still missing due to the early stage of the project. The preliminary results are discussed in the next section.

\section{Preliminary Results}
The first results are promising. The implementation of the node is feasible where the only hard task is to ensure that the security components are able to access the operating system. The installation of the security components in the security environment installed in a virtual machine is also challenging because this influences the performance and the security of the security system. 

Security issues of this organisation are analysed and they are solvable when up-to-date approaches from cryptography are used. Especially the more and more emerging integrity checks and the complete installation of a distributed public/private key infrastructure are important. With this, an adversarial is not able to use the distributed security system for attacking the network.

\subsection{Infection Handling}
One main advantage of the proposed architecture is the infection handling. If some security component identifies a virtual machine of a node as infected, the following workflow is processed: the security environment of the node halts the infected virtual machine to prevent propagation. It duplicates the virtual machine and sends this to the security server to analyse it more deeply and to save the evidences for legal aspects - this is a weakness in current systems: either the node is disinfected and the evidences are destroyed or the evidences are saved but the node is still infected. Afterwards, the security server provides a clean virtual machine with the desired applications installed in order to limit the downtime of the node.

\subsection{Security Issues}
The proposed architecture is used to implement a distributed security system with integrated components. The security components are able to roam through the network and cooperative workflows enable the detection of unforeseen intrusions. This has a major drawback that adversaries may use the system to attack a network, i.e. to propagate intrusions through it. This must be prevented through the design of the architecture and is discussed now.

The layer one containing the hardware is not an aim of adversaries. Layer two with the core operating and virtualization system is furthermore highly dependent on the facilitated hardware and changes therefore only when the hardware changes. This is ensured through public/private key signatures used in cryptography (this is also integrity checking called). When an adversarial installs an intrusions in this layer, it is immediately recognised and prevented. Layer three and four contains the operating system with application of the user, which are protected as the operating system in nowadays implementations. The virtual machine containing the security environment and the security components is additionally secured: the security environment does not change and is therefore protected using a cryptographic signature with integrity checking. The security components access resources of the node. These are secured using public/private keys organised in a distributed public key infrastructure. Only when the security components have the right keys, they receive access to the resources where security components initiated by the adversarial does not have these. To summarise the security issues, the adversarial is still able to install intrusions in the operating system and in the security system when the implementation provides bugs or when the adversarial knows internal knowledge. 

\section{Conclusion}
\label{secConclusion}
This article discusses how the features of virtualization systems with service oriented architecture increase the performance of a network security system. The advantages especially help to identify novel more and more intelligent intrusions and provides a more sophisticated infection handling. Furthermore, the article faces several unsolved problems, which are of interested for novel network security systems. The next step in the project is to implement a prototype of the architecture and to build up a testbed with some nodes to simulate the whole workflow. For this, a VMware implementation is first desired due to the reduced time for setting up all of these nodes. The architecture is also usable in current approaches of network security systems facilitating artificial immune systems \cite{Hil06b}, multi-agent systems, and distributed systems to distinguish the normal operating system and the security system on each node \cite{Spa00}.

\section*{Acknowledgments}
This work is supported by the Ministre Luxembourgeois de l'education et de la recherche and the University of Luxembourg. 

\bibliography{paper}
\bibliographystyle{plain}

\end{document}